\documentclass[twocolumn,preprint]{elsart3p}

\usepackage{epsfig}

\usepackage{amssymb}

\begin{document}

\begin{frontmatter}
{\hfill \bf Published in Phys. Lett. B}
\title{Probing the critical behavior in the evolution of GDR width at very low temperatures in A $\sim$ 100 mass region}
\author[label1]{Balaram Dey},
\author[label1]{Debasish Mondal},
\author[label1]{Deepak Pandit},
\author[label1]{S. Mukhopadhyay},
\author[label1]{Surajit Pal},
\author[label2]{Srijit Bhattacharya},
\author[label3]{A. De},
\author[label1]{K. Banerjee},
\author[label4]{N. Dinh Dang},
\author[label5]{N. Quang Hung},
\author {and}
\author[label1]{S. R. Banerjee\corauthref{cor}}
\corauth[cor]{Corresponding author.}
\ead{srb@vecc.gov.in}

\address[label1]{Variable Energy Cyclotron Centre, 1/AF-Bidhannagar, Kolkata-700064, India}
\address[label2]{Department of Physics, Barasat Govt. College, Barasat, N 24 Pgs, Kolkata - 700124, India }
\address[label3]{Department of Physics, Raniganj Girls' College, Raniganj - 713358, India}
\address[label4]{Theoretical Nuclear Physics Laboratory, RIKEN Nishina Center for Accelerator-Based Science, RIKEN, 2-1 Hirosawa, Wako city, Saitama 351-0198, Japan}
\address[label5]{School of Engineering, Tan Tao University, Tan Tao University Avenue, Tan Duc Ecity, Duc Hoa, Long An Province, Viet Nam}

\begin{abstract}
The influence of giant dipole resonance (GDR) induced quadrupole moment on GDR width at low temperatures is investigated experimentally by measuring GDR width systematically in the unexplored temperature range $T$=0.8-1.5 MeV, for the first time, in $A$ $\sim$ 100 mass region. The measured GDR widths, using alpha induced fusion reaction, for $^{97}$Tc confirm that the GDR width remains constant at the ground state value up to a critical temperature and increases sharply thereafter with increase in $T$. The data have been compared with the adiabatic Thermal Shape Fluctuation Model (TSFM), phenomenological Critical Temperature Fluctuation Model (CTFM) and microscopic Phonon Damping Model (PDM). Interestingly, CTFM and PDM give similar results and agree with the data, whereas the TSFM differs significantly even after incorporating the shell effects.
\end{abstract}

\end{frontmatter}

One of the fascinating areas of experimental nuclear physics has been the study of the giant dipole resonance (GDR) built on the excited states of atomic nuclei. These experimental studies, over the years, have shown that the GDR width increases with both temperature $T$ and angular momentum $J$, whereas its centroid energy remains mostly unchanged as $T$ and $J$ vary \cite{hara01,gaardhoje}. It is worth mentioning that the effect of $J$ and $T$ on GDR width becomes noticeable only above a critical angular momentum, $J_c$$\sim$0.6$A^{5/6}$ \cite{kusn01} and $T$ $\approx$ 1 MeV. Although, a wealth of data exists on the angular momentum dependence of GDR width in different mass regions \cite{drebi,braco01,cam01,mat,srijit01,drc,mukul}, the measurement of the GDR width at low temperatures ($T$ $<$ 1 MeV) is rather scarce due to the experimental difficulties in populating the nuclei at low excitation energies. 
The present work aims at providing systematic experimental data on GDR width  at this very low temperature region. It is also our endeavor to systematically assess different theoretical models and understand the complete nature of the damping mechanism as a function of $T$ inside the atomic nucleus.

A number of theoretical approaches have been proposed to demonstrate the behavior of GDR width as a function of $T$ and $J$. Microscopically, the increase of GDR width as a function of $T$ is described reasonably well within the Phonon Damping Model (PDM) \cite{dang01,dang02,dang03}. 
The PDM calculates the GDR width and the strength function directly in the laboratory frame without any need for an explicit inclusion of thermal fluctuation of nuclear shapes. Interestingly, the macroscopic Thermal Shape Fluctuation Model (TSFM) \cite{gal,pach,good,dub,alh88,ormand,maj01,dipu1}, on the other hand, is based on the fact that large-amplitude thermal fluctuations of nuclear shape play an important role in describing the increase of GDR width as a function of $T$.
This model explains very well the $J$ dependence of the GDR width, the mass dependence of the critical angular momentum ($J_c$) and the Jacobi shape transition \cite{maj01,dipu1}. However, it is unable to explain the $T$ dependence below 1.5 MeV in different mass regions \cite{heck01,cam02,dipu3,supm1,dipu4}.
Recently, a new model has been proposed by modifying the phenomenological parameterization (pTSFM) \cite{kusn01} based on the TSFM and is called the Critical Temperature included Fluctuation Model (CTFM) \cite{dipu4}. The CTFM provides a good description of the behavior of GDR width for both $T$ and $J$ in the entire mass region \cite{dipu4,dipu2,dipu6}. Unfortunately, the number of GDR width measurements till now at $T$ $<$ 1 MeV are inadequate to test either the critical behavior of the GDR width or to conclude that the GDR width remains nearly constant at the ground state value below $T$ $\sim$ 1 MeV as predicted by both PDM and CTFM.

\begin{figure}
\begin{center}
\includegraphics[height=5.7 cm, width=5.5 cm]{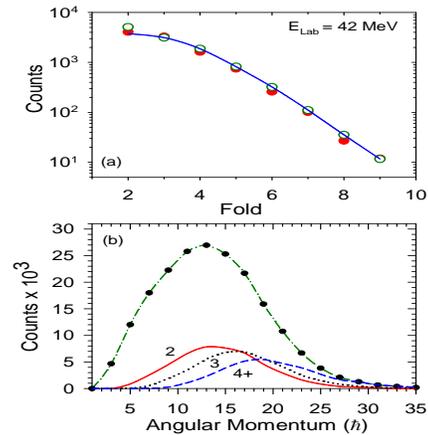}
\caption{\label{nb_ang} [Color online] [Top panel] Measured fold distributions with high energy $\gamma$-rays (filled circles) and with neutrons (open circles) along with the GEANT4 simulation. [Bottom panel] Angular momentum distributions for different folds at 42 MeV incident energy along with the incident distribution (dot-dashed with symbols).}
\end{center}
\end{figure}
In order to examine the above queries regarding the behavior of  GDR apparent width as function of $T$, a systematic measurement of  GDR apparent width in the unexplored region ($T$ $=$ 0.8 $-$ 1.5 MeV) was performed for $^{97}$Tc using alpha induced fusion reactions. This is the first measurement of GDR width at finite temperature in the $A$ $\sim$ 100 mass region. The measured GDR apparent widths, both above and below the critical point, can be effectively used to verify the existing theoretical models.

The experiments were performed at the Variable Energy Cyclotron Centre (VECC), Kolkata. A self supporting 1 mg/cm$^2$ thick $^{93}$Nb target was bombarded with alpha beams produced by the K-130 cyclotron. Four different beam energies of 28, 35, 42 and 50 MeV were used to form the compound nucleus (CN) $^{97}$Tc at the excitation energies of 29.3, 36, 43 and 50.4 MeV, respectively. The high energy $\gamma$-rays from the decay of $^{97}$Tc were detected using the high energy photon spectrometer LAMBDA \cite{supm2}. A part of the spectrometer consisting of 49 BaF$_2$ detectors (each having dimension of 3.5$\times$3.5$\times$35 cm$^3$), was arranged in a 7$\times$7 matrix configuration. The spectrometer was placed at a distance of 50 cm from the target (covering 1.8$\%$ of 4$\pi$) and at an angle of 90$^{\circ}$ with the beam axis. Along with the LAMBDA spectrometer, a 50 element low energy $\gamma$-multiplicity filter \cite{dipu5} was also used to estimate the angular momentum populated in the CN as well as to get a fast start trigger for time-of-flight (TOF) measurement. The multiplicity filter was split into two blocks of 25 detectors each, in a staggered castle type geometry to equalize the solid angle for each multiplicity detector element, and placed at a distance of 5 cm above and below the centre of the target. The efficiency of the multiplicity set-up was 56$\%$ as calculated using GEANT4 \cite{geant4} simulation. The TOF technique was used to separate neutron background in the high-energy $\gamma$-spectrum. Pile up events were rejected using the pulse shape discrimination (PSD) technique by measuring the charge deposition over two integrating time intervals (50 ns and 2 $\mu$s) in each of the detectors.

\begin{figure}
\begin{center}
\includegraphics[height=7.9 cm, width=6.5 cm]{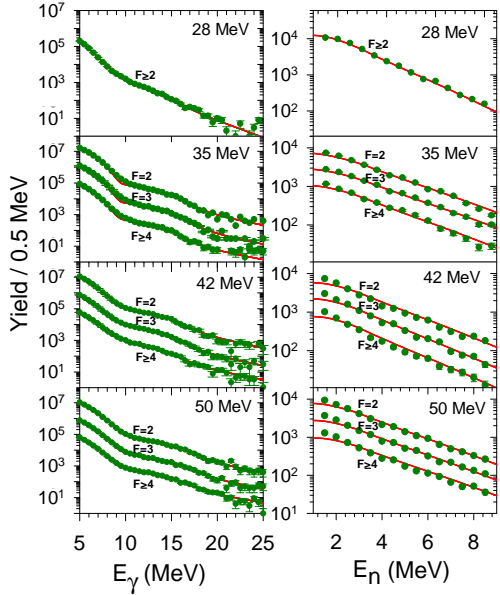}
\caption{\label{n-gamma}
[Color online] The high energy $\gamma$-ray spectra (filled circles) along with the CASCADE predictions (continuous line)
for different folds (F) [left panel] and the neutron evaporation energy spectra (filled circles) along with the CASCADE predictions (continuous line) for different folds (F) [right panel] at incident energies of 28, 35, 42 and 50 MeV. $F$=2 and $F$=3 data have been multiplied by 100 and 10, respectively.}
\end{center}
\end{figure}

The experimentally measured fold distributions were converted into angular momentum distributions using a realistic technique \cite{dipu5} based on GEANT4 
simulation. The measured fold distribution for the reaction $^4$He + $^{93}$Nb at 42 MeV incident energy is shown in Fig \ref{nb_ang}a. The extracted angular momentum distributions corresponding to different folds have also been shown in Fig \ref{nb_ang}b. 
Recently, the inverse level density parameter ($k$) was extracted from evaporated neutrons, protons and alpha particles \cite{pra} from the same system at 35 MeV. It was observed that the absolute values of $k$ obtained from different particle spectra were different but in all cases the value of $k$ decreased with increase in angular momentum. Hence, to fix the inverse level density parameter, the evaporated neutron spectrum was also measured independently by employing a liquid organic scintillator (BC501A) based neutron detector \cite{kban1} in coincidence with the $\gamma$- multiplicities. The neutron detector was placed at a distance of 1.5 m from the target position at an angle of 90$^{\circ}$ with respect to the beam axis. The neutron TOF spectra were converted to energy spectra by considering the prompt $\gamma$-peak as time reference. Efficiency correction for the neutron detector was carried out using GEANT4 simulation \cite{geant4}. The evaporated neutron energy spectra corresponding to different folds were compared with the CASCADE calculation \cite{cas} to determine the nuclear level density (NLD) parameter using a chi-square minimization technique in the energy range of 2-7 MeV (Fig \ref{n-gamma}, right panel). The fold distribution measured in coincidence with neutrons is compared with the fold distribution obtained in coincidence with  high energy $\gamma$- rays for 42 MeV incident energy (as shown in Fig \ref{nb_ang}a). The good match between the two fold distributions indicates that the populated spin distributions in both cases are similar.

\begin{figure}
\begin{center}
\includegraphics[height=7.9 cm, width=6.5 cm]{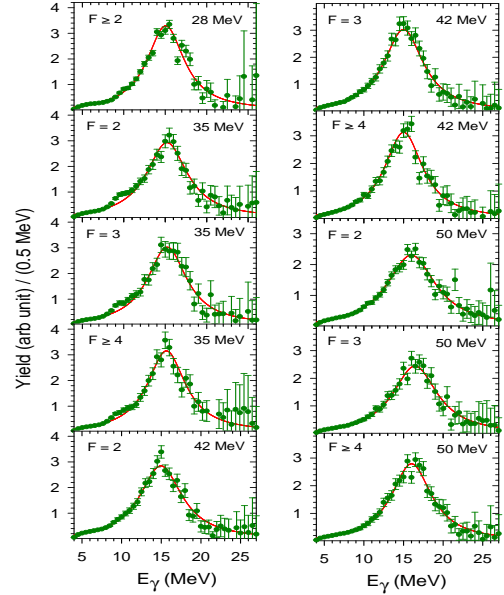}
\caption{\label{divided} [Color online] 
Linearized GDR plots are shown (symbols) using the quantity $F(E_{\gamma}$)$Y^{exp}$($E_{\gamma}$)/$Y^{cal}$($E_{\gamma}$), where $Y^{exp}$($E_{\gamma}$) and $Y^{cal}$($E_{\gamma}$) are the experimental and best fitted CASCADE spectra, respectively, corresponding to the single Lorentzian function $F(E_{\gamma}$) used in the CASCADE (continuous line).}
\end{center}
\end{figure}

\begin{figure}
\begin{center}
\includegraphics[height=4.0 cm, width=7.0 cm]{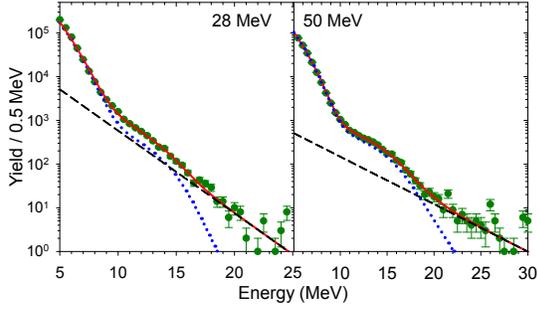}
\caption{\label{brem} [Color online] The experimental $\gamma$-ray energy spectra (symbols) at 28  and 50 MeV are compared with the CASCADE prediction along with the bremsstrahlung component (continuous line). The individual CASCADE (dotted) and bremsstrahlung (dashed) component are also shown.}
\end{center}
\end{figure}

The high energy $\gamma$-ray spectra for different folds of the multiplicity filter (Fig \ref{n-gamma}, left panel) were extracted in offline analysis using the cluster summing technique \cite{supm2}. GDR widths were obtained from the measured high energy $\gamma$-ray spectra by comparing it with the statistical model calculation CASCADE along with a bremsstrahlung component. 
Bremsstrahlung emission was parameterized by the exponential function ($e^{-E_{\gamma}/E_0}$). $E_0$ was adopted from the systematic $E_0$ = 1.1[($E_{Lab}$ - $V_{c}$)/$A_{p}$]$^{0.72}$ \cite{nifnecker}, where $E_{Lab}$, $V_{c}$ and $A_{p}$ represent the beam energy, Coulomb barrier and projectile mass, respectively. The systematic was verified earlier \cite{supm1} for alpha beams at similar energies by measuring the angular distribution of $\gamma$-rays arising from the non-statistical component. The CASCADE calculation as well as the bremsstrahlung component (both folded with the detector response function) are shown in Fig \ref{brem} along with the experimental data at 28  and 50 MeV incident energies. The response function of the LAMBDA spectrometer was generated using GEANT4 simulation. 
In the statistical model calculation, a single Lorentzian GDR strength function was assumed, having centroid energy ($E_{GDR}$) and width ($\Gamma$) as parameters. The other parameters were kept fixed as used for describing the neutron evaporation spectra. The moment of inertia of the CN was taken as $I$ = $I_0$(1 + $\delta_1J^2$ + $\delta_2J^4$), where $I_0$ is the moment of inertia of the spherical nucleus. The parameters $r_0$, $\delta_1$ and $\delta_2$ were kept at their default values of 1.2 fm, 0.9 $\times$ 10$^{-5}$ and 2.0 $\times$ 10$^{-8}$, respectively.
The level density prescription of Ignatyuk \cite{igna} was taken with the asymptotic level density parameters as extracted from the corresponding neutron evaporation spectra. The simulated spin distributions deduced from the experimental fold distributions were used as inputs for different folds for both neutron and high energy $\gamma$-ray analyses. The GDR widths were obtained from the best fit statistical model calculations using a $\chi^2$ minimization (in the energy range of 10-20 MeV). In order to highlight the GDR region, linearized GDR plots are shown in Fig \ref{divided} using the quantity $F(E_{\gamma}$)$Y^{exp}$($E_{\gamma}$)/$Y^{cal}$($E_{\gamma}$), where $Y^{exp}$($E_{\gamma}$) and $Y^{cal}$($E_{\gamma}$) are the experimental and best fitted CASCADE spectra, respectively, corresponding to the single Lorentzian function $F(E_{\gamma}$).

As $\gamma$-rays from the GDR are emitted from various steps of the compound nuclear decay chain, the average values of $J$ and $T$ have been considered. 
While estimating the average temperature, a lower limit in the excitation energy ($E^*$) during the CN decay process was employed in the statistical model calculation in accordance with the prescription described in Ref \cite{wieland,srijit02}. This lower limit in $E^*$ is selected when the cut off in the excitation energy only affects the $\gamma$-emission at very low energy but does not alter the GDR region. The average value of $T$ was calculated from $\overline{E}^*$ using the relation $\overline{T}$ = [($\overline{E}^*$ - $\overline{E}_{rot}$ - $E_{GDR}$ -$E_p$)/a($\overline{E}^*$)]$^{1/2}$, where $a(\overline{E}^*$) is the excitation energy-dependent level density parameter and $E_p$ is the pairing energy. $\overline{E}_{rot}$ was computed at $\overline{J}$ within the CASCADE corresponding to each fold. $\overline{E}^*$ is the average of the excitation energy weighted over the daughter nuclei for $\gamma$-emission in the GDR region from $E_{\gamma}$ = 10 - 20 MeV given as $\overline{E}^*$ = $\sum$E$^{*}_{i}$w$_i$/$\sum$w$_i$. $E^{*}_{i}$ is the excitation energy of $i^{th}$ nuclei in the decay steps and $w_i$ is the yield in the region $E_{\gamma}$ = 10 - 20 MeV. The extracted GDR parameters, $\overline{T}$ and $\overline{J}$ are given in Table I.
The error estimation of temperature includes the uncertainty in the level density parameter, the effect of varying GDR centroid energy and the width of the selected angular momentum distribution. It needs to be mentioned that nuclear deformation was not included in our statistical calculation. Hence, we report on the extraction of the GDR apparent widths and compare them with the different models, which also provide the apparent width of the GDR.

\begin{figure}
\begin{center}
\includegraphics[height=8 cm, width=6.0 cm]{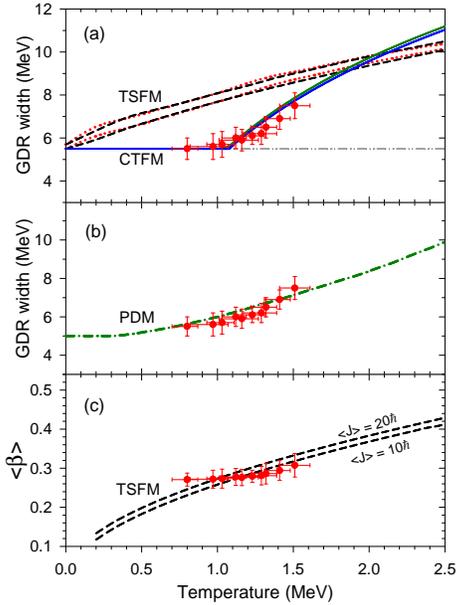}
\caption{\label{width} [a] GDR width as a function of temperature. Experimental data (symbols) have been compared with the TSFM calculations with shell effect (dotted lines) and without shell effect (dashed lines) are shown for $\overline{J}$ = 0$\hbar$ (lower) and $\overline{J}$ = 30$\hbar$ (upper). The CTFM predictions for $\overline{J}$ = 10$\hbar$ (lower) and $\overline{J}$ = 20$\hbar$ (upper) are shown with continuous lines. The double dot-dashed line represents the ground state value of the GDR width. [b] The PDM prediction is shown by the dotted-dashed. [c] The empirical deformations (symbols) extracted from the experimental GDR widths compared with the TSFM predictions (dashed lines) at two angular momenta.}
\end{center}
\end{figure}

The GDR widths measured in the present work at low $T$ (0.8 - 1.5 MeV) are shown in Fig \ref{width}a. The data have been compared with the theoretical predictions based on the TSFM. Within this model, the GDR strength function is calculated by averaging the lineshapes corresponding to the different possible deformations of the nuclear shape. The averaging over the distribution of shapes is weighted with a Boltzmann factor $e^{-F(\beta,\gamma)/T}$, where $F(\beta$,$\gamma$) is the free energy and $T$ is the nuclear temperature \cite{gal,pach,good,dub,alh88,ormand,maj01,dipu1}. The calculations were performed with (dotted) and without (dashed) considering the shell effect \cite{stru1,brac}(Fig \ref{width}a). As expected for $^{97}$Tc, the effect of shell correction on the GDR width is quite small and leads
to similar results as obtained considering the liquid drop model. The TSFM calculations also show that the effect of angular momentum on the GDR width below 30 $\hbar$ is small and essentially remains unchanged below 20 $\hbar$. The compound nucleus particle evaporation widths ($\Gamma$$_{ev}$) have been incorporated in the TSFM calculation to take into consideration the effect of evaporation of particles and the corresponding energy loss before the GDR $\gamma$-emission in the CN decay chain. In this low temperature region, the particle decay width is rather small ($\sim$ 0.2 MeV at $T$ = 2 MeV) and its inclusion within the TSFM marginally improves the prediction. The predictions of TSFM at $J$ = 0 and $J$ = 30 $\hbar$ are shown in Fig \ref{width}a and compared with the experimental data. As can be seen, the adiabatic TSFM differs significantly from the experimental data in the entire region. This interesting result clearly suggests that the suppression of the GDR width at these low $T$ is a general feature for all nuclei and cannot be explained including only shell effects.
It is also very interesting and important point to note that the GDR width data indeed remains nearly constant till $T$ $\sim$ 1 MeV (Fig \ref{width}a) and increases thereafter as predicted by PDM and CTFM. In order to compare the data with CTFM, the ground state GDR width was calculated using the ground state deformation ($\beta$ = 0.134) \cite{moller} and spreading width parameterization $\Gamma_s$ = 0.05$E_{GDR}^{1.6}$ \cite{jungh01} for each Lorentzian. The ground state value was estimated to be 5.5 MeV which is consistent with the experimentally measured value in this mass region \cite{carlos}. The behavior of the GDR width within the CTFM was calculated as a function of $T$  for $J$ = 10 and 20 $\hbar$ selecting the extreme angular momenta involving the experimental data.
The $\Gamma_{ev}$ was not included in the CTFM calculations as the model was put forward by fitting the experimental data. Interestingly, the CTFM represents the data remarkably well over the entire $T$ region. This excellent match between the experimental data and the CTFM clearly suggests that the experimental GDR widths are not suppressed, rather TSFM over predicts the GDR width at low temperatures as it does not take into account the intrinsic GDR fluctuations induced by the GDR vibrations. Moreover, the systematic trend of the data also shows that the critical temperature for the increase of GDR width is between 1 and 1.2 MeV as predicted by the CTFM ($T_c$ = 0.7 + 37.5/$A$).

The data were also compared with the results of microscopic PDM calculations \cite{dang01,dang02,dang03}.  Within the PDM, the GDR damping mechanism is caused by coupling of the GDR to noncollective particle-hole (ph) and particle-particle (pp) [hole-hole (hh)] configurations. The coupling to  the various ph configurations leads to the quantal width (exists even at $T$ = 0), whereas the thermal width arises owing to the coupling to pp and hh configurations, which appear at $T$ $>$ 0 because of the distortion of the Fermi surface. The model emphasizes the inclusion of thermal pairing, since, in finite systems such as atomic nuclei, thermal pairing does not collapse at the critical temperature $T_{cp} = 0.57\Delta(T=0)$ of the superfluid-normal phase transition in infinite systems, but decreases monotonically as $T$ increases. The prediction of the PDM is shown in Fig. \ref{width}b. The calculations were performed at $J=$ 0 by using the single-particle energies obtained within the deformed Woods-Saxon potentials with the deformation parameter $\beta$ = 0.134, and including exact canonical-ensemble thermal pairing gaps for neutrons and protons~\cite{dang03}. 
As can be seen in Fig. \ref{width}b, the PDM describes the data quite well in the entire $T$ range using a width of around 5 MeV at $T=0$, which is close to the deformed ground state GDR width. It is intriguing to find that, even though the formalisms of PDM and CTFM seem to be completely different in origin, the two models give very similar results. 
It would also be interesting to compare the data with TSFM by including the effect of thermal pairing, but is beyond the scope of this present work. Nevertheless, the present experimental study does provide a stringent testing ground of the theoretical models as a function of $T$. 
The average deformation ($<$$\beta$$>$) for this case was also extracted using the universal correlation between the experimental GDR width and the average deformation of the nucleus at finite $T$, and was compared with the TSFM. The correlation has been proposed recently by including the deformation induced by the GDR motion \cite{dipu2}. As can be observed from Fig \ref{width}c, the empirical deformations extracted from the experimental data match excellently well with the TSFM calculation above $T_c$. The good description of the CTFM as well as the validity of the universal correlation indicate that GDR induced deformations could play a decisive role in suppressing the GDR width at low $T$.


\begin{table}
\caption{\label{data} Average temperatures and average angular momenta along with level density parameters, GDR widths, centroid energies and bremsstrahlung parameters at different beam energies.}  
\vspace{5mm}

\begin{center}		
\begin{tabular}{|c|c|c|c|c|c|c|c|}
\hline

$E_{lab}$  & $E^{*}$  &$\;$$\overline{J}$$\;$&$\overline{T}$&$\;$$A/\ \widetilde{a}$$\;$& $\Gamma_{GDR}$ & $E_{GDR}$ & $E_0$  \\
MeV & MeV      &$\hbar$& MeV & MeV & MeV & MeV & MeV \\ \hline
28 & 29.3     & 13$\pm$6&$\;$0.80$^{+0.07}_{-0.10}$$\;$& 8.0$\pm$0.4 & 5.5$\pm$0.5& 15.2$\pm$0.1 & 2.4  \\  \hline   
   &      & 13$\pm$4&$\;$1.12$^{+0.07}_{-0.09}$$\;$& 9.7$\pm$0.5 & 6.0$\pm$0.5& 15.5$\pm$0.1 & 3.4 \\
35 & 36.0     & 15$\pm$5&$\;$1.03$^{+0.08}_{-0.10}$$\;$& 9.5$\pm$0.3 & 5.7$\pm$0.6& 15.5$\pm$0.1 & 3.4  \\
   &      & 18$\pm$5&$\;$0.97$^{+0.10}_{-0.15}$$\;$& 8.2$\pm$0.4 & 5.6$\pm$0.6& 15.4$\pm$0.1 & 3.4 \\ \hline
   &      & 14$\pm$5&$\;$1.32$^{+0.07}_{-0.10}$$\;$& 9.0$\pm$0.4 & 6.5$\pm$0.5& 15.3$\pm$0.1 & 4.2 \\
42 & 43.0     & 16$\pm$5&$\;$1.23$^{+0.07}_{-0.10}$$\;$& 8.1$\pm$0.4 & 6.1$\pm$0.4& 15.3$\pm$0.1 & 4.2 \\
   &      & 19$\pm$6&$\;$1.16$^{+0.11}_{-0.15}$$\;$& 7.8$\pm$0.5 & 5.9$\pm$0.5& 15.3$\pm$0.1 & 4.2 \\ \hline
   &      & 14$\pm$5&$\;$1.51$^{+0.09}_{-0.09}$$\;$& 9.2$\pm$0.5 & 7.5$\pm$0.6& 16.4$\pm$0.1 & 4.8 \\
50 & 50.4     & 16$\pm$5&$\;$1.41$^{+0.07}_{-0.10}$$\;$& 8.5$\pm$0.4 & 6.9$\pm$0.5& 16.6$\pm$0.1 & 4.8 \\
   &      & 20$\pm$5&$\;$1.29$^{+0.09}_{-0.12}$$\;$& 8.2$\pm$0.4 & 6.2$\pm$0.5& 16.3$\pm$0.1 & 4.8 \\ \hline

\hline
\end{tabular}
\end{center}		
\end{table}


In summary, a systematic study of the GDR width as a function of $T$ has been presented in the unexplored region ($T$=0.8-1.5 MeV) for $^{97}$Tc. In order to determine the temperature precisely, the level density parameter has been extracted from the neutron evaporation spectrum and the angular momentum from gamma multiplicity filter using a realistic approach. The systematic trend of the data shows that GDR width remains nearly constant at the ground state value up to $T$ $\sim$ 1 MeV and increases thereafter. The microscopic PDM and phenomenological CTFM describe the data reasonably well, whereas the adiabatic TSFM differs substantially even after inclusion of shell effect. These interesting results indicate that the effect of GDR induced deformation could be one of the ways in explaining macroscopically the behavior of GDR width at low $T$. However, this effect is not explicitly needed in microscopic PDM, rather thermal pairing should be included to have adequate description of the damping of GDR width in open shell nuclei at low $T$.  



\section*{Acknowledgements}
The authors are thankful to VECC Cyclotron operators for a smooth running of the accelerator during the experiment. N. Quang Hung acknowledges the support by the National Foundation for Science and Technology Development (NAFOSTED) of Vietnam through Grant No. 103.04-2013.08.


\begin{thebibliography}{99}

\bibitem{hara01} M. N. Harakeh and A. van der Woude Giant Resonances, Fundamental High-frequency Modes of
Nuclear Excitation, Clarendon Press, Oxford, 2001.
\bibitem{gaardhoje} J. J. Gaardhoje Annu. rev. Nucl. Part. Sci. 42 (1992) 483.

\bibitem{kusn01} D. Kusnezov et al., Phys. Rev. Lett. 81 (1998) 542.

\bibitem{drebi} Z. M. Drebi et al., Phys. Rev. C 52 (1995) 578.
\bibitem{braco01} A. Bracco et al., Phys. Rev. Lett. 74 (1995) 3748.
\bibitem{cam01} F. Camera et al., Phys. Rev. C 60 (1999) 014306.
\bibitem{mat} M. Mattiuzzi et al., Phys. Lett. B 364 (1995) 13.
\bibitem{srijit01} Srijit Bhattacharya et al., Phys. Rev. C 77 (2008) 024318.
\bibitem{drc} D. R. Chakrabarty et al., J. Phys. G: Nucl. Part. Phys. 37 (2010) 055105.
\bibitem{mukul} I. Mukul et al., Phys. Rev. C 88 (2013) 024312.


\bibitem{dang01} N. Dinh Dang  and A. Arima, Phys. Rev. Lett. 80 (1998) 4145, Nucl. Phys. A 636 (1998) 427.
\bibitem{dang02} N. Dinh Dang, Phys. Rev. C 85 (2012) 2023.
\bibitem{dang03} N. Dinh Dang and N. Quang Hung, Phys. Rev. C  86 (2012) 044333.

\bibitem{gal} M. Gallardo et al., Phys. Lett. B 191 (1987) 222.
\bibitem{pach} J. M. Pacheco et al., Phys. Rev. Lett. 61 (1988) 294.
\bibitem{good} A. L. Goodman, Nucl. Phys. A 528 (1991) 348.
\bibitem{dub} N. Dubray et al., Acta. Phys. Pol B 36 (2005) 1161.
\bibitem{alh88} Y. Alhassid et al., Phys. Rev. Lett. 61 (1988) 1926.
\bibitem{ormand} W. E. Ormand et al., Phys. Rev. Lett. 77 (1996) 607.

\bibitem{maj01} A. Maj et al., Nucl. Phys. A 731 (2004) 319.
\bibitem{dipu1}  Deepak Pandit et al., Phys. Rev. C 81 (2010) 061302(R).

\bibitem{heck01} P. Heckman et al., Phys. Lett. B 555 (2003) 43.
\bibitem{cam02} F. Camera et al., Phys. Lett. B 560 (2003) 155.
\bibitem{dipu3}  Deepak Pandit et al., Phys. Lett. B 690 (2010) 473.
\bibitem{supm1} S. Mukhopadhayay et al., Phys. Lett. B 709 (2012) 9.
\bibitem{dipu4} Deepak Pandit et al., Phys. Lett. B 713 (2012) 434.

\bibitem{dipu2}  Deepak Pandit et al., Phys. Rev. C 87 (2013) 044325.
\bibitem{dipu6}  Deepak Pandit et al., Phys. Rev. C 88 (2013) 054327.

\bibitem{supm2} S. Mukhopadhayay et al., Nucl. Instr. and Meth. A 582 (2007) 603.	 
\bibitem{dipu5} Deepak Pandit et al., Nucl. Instr. and Meth. A 624 (2010) 148.
\bibitem{geant4} S. Agostinelli, et al., Nucl. Instr. and Meth. A 506 (2003) 250.
\bibitem{pra} Pratap Roy, et al., Phys. Rev. C 86 (2012) 044622.

\bibitem{kban1} K. Banerjee et al., Nucl. Instr. and Meth. A 608 (2009) 440.

\bibitem{cas} F. Puhlhofer, Nucl. Phys. 280 (1977) 267.
\bibitem{nifnecker} H. Nifennecker and J. A. Pinston, Annu. Rev. Nucl. Part. Sci. 40 (1990) 113.
\bibitem{igna} A. V. Ignatyuk, G. N. Smirenkin, and A. S. Tishin, Sov. J. Nucl. Phys. 21 (1975) 255 [Yad. Fiz. 21 (1975) 485].

\bibitem{wieland} O. Wieland et al., Phys. Rev. Lett. 97, (2006) 012501.
\bibitem{srijit02} Srijit Bhattacharya et al., Phys. Rev. C  78 (2008) 064601.

\bibitem{stru1} V. M. Strutinsky, Nucl. Phys. A 95 (1967) 420, Nucl. Phys. A 122 (1968) 1.
\bibitem{brac} M. Brach and P. Quentin et al., Nucl. Phys. A 361 (1969) 35.


\bibitem{moller} P. Moller et al., At. Data Nucl. Data Tables 59 (1995) 185.
\bibitem{jungh01}  A. R. Junghans et al., Phys. Lett. B 670 (2008) 200.
\bibitem{carlos} P. Carlos et al., Nucl. Phys. A 219 (1974) 61.



 









\end{thebibliography}
\end{document}